\documentclass[%
 jap,
 jmp,%
 amsmath,amssymb,
 reprint,%
]{revtex4-1}

\usepackage{graphicx}
\usepackage{dcolumn}
\usepackage{bm}
\usepackage{amsmath}
\usepackage{ amssymb }

\begin{document}





\title{Detecting nonlinear acoustic waves in liquids with nonlinear dipole optical antennae}
\author{Ivan S. Maksymov}
\email{ivan.maksymov@rmit.edu.au} 
\author{Andrew D. Greentree}
\affiliation{ARC Centre of Excellence for Nanoscale BioPhotonics, School of Applied Sciences, RMIT University, Melbourne, VIC 3001, Australia}

\date{\today}

\begin{abstract}
Ultrasound is an important imaging modality for biological systems. High-frequency ultrasound can also (e.g., via acoustical nonlinearities) be used to provide deeply penetrating and high-resolution imaging of vascular structure via catheterisation. The latter is an important diagnostic in vascular health. Typically, ultrasound requires sources and transducers that are greater than, or of order the same size as the wavelength of the acoustic wave. Here we design and theoretically demonstrate that single silver nanorods, acting as optical nonlinear dipole antennae, can be used to detect ultrasound via Brillouin light scattering from linear and nonlinear acoustic waves propagating in bulk water. The nanorods are tuned to operate on high-order plasmon modes in contrast to the usual approach of using fundamental plasmon resonances.  The high-order operation also gives rise to enhanced optical third-harmonic generation, which provides an important method for exciting the higher-order Fabry-Perot modes of the dipole antenna.
  
\end{abstract}


\maketitle 

\section{Introduction}

Control of light at the nanoscale is a burgeoning field, and optical antennae are fast emerging as one of the most important tools for collecting, emitting and more generally controlling light with nanoscale (sub wavelength) elements \cite{Nov11, Kra13}.  This functionality derives from strong coupling of optical fields to plasmon modes in the antennae, where the resonance of the plasmon modes can be controlled by the geometry of the nanoscale elements.

We are interested in the detection of high-frequency $\sim 20-150$ MHz ultrasound used for intravascular photoacoustic (IVPA) imaging \cite{Wan10}. IVPA imaging is complementary to traditional intravascular ultrasound (IVUS) imaging \cite{Wan10, Mae09, Ma15}. In IVUS imaging, ultrasound is used to interrogate the body, and an image is reconstructed from acoustic waves that are backscattered from blood vessel tissues. In IVPA imaging, a laser pulse is used to interrogate the body instead of ultrasound, and an image is reconstructed from broadband acoustic waves generated by light absorption events in various tissue components \cite{Wan10}. Consequently, IVPA imaging is less affected by the light scattering in tissues, and therefore, can penetrate deeper as compared with purely optical imaging methods \cite{Wan10, Mae09, Ma15}. Moreover, the contrast of IVUS and IVPA imaging can be increased by exploiting harmonics of ultrasound generated due to backscattering from nonlinear contrast agents \cite{Goe06}. This fact additionally motivates our interest in the nonlinear acoustic interaction and optical detection of harmonics of ultrasound. 

Contrast in PA imaging arises from the natural variation in the optical absorption of tissue components. The absorption cross-section of an optical antenna is many orders of magnitude higher than that of tissues.  Consequently, recent attention has turned to the use of optical antennae (plasmonic nanoparticles) as contrast agents for IVPA imaging \cite{Hom12} and all-optical sources of high-frequency ultrasound \cite{Hou06}. 

Hereafter we theoretically demonstrate that in addition to their ability to generate high-frequency ultrasound, single optical antennae immersed into a bulk liquid can also be used to detect ultrasound. Ultrasonic transduction is via the detection of weak optical signals arising due to the Brillouin light scattering (BLS) from acoustic waves propagating in the liquid. This transduction mechanism is different from that proposed in Ref.~\onlinecite{Yak13}, where a large array of plasmonic metamaterials was used for non-resonant optical detection of ultrasound.

\begin{figure}[htb]
\centering\includegraphics[width=8.5cm]{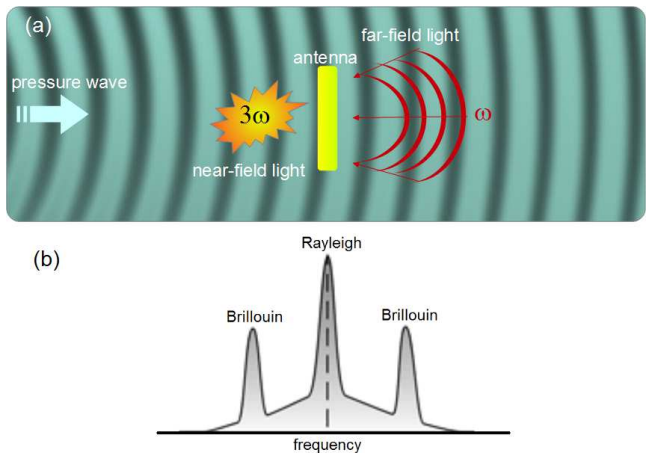}
\caption{(a) Schematic of the detection of acoustic waves using an optical dipole antenna immersed into a liquid. An acoustic wave propagating in the liquid (blue background) is probed by an optical signal localised in the near-field-zone of the antenna (yellow rectangle). This signal originates from the nonlinear generation of the third harmonic of the frequency of the freely propagating light focused on the dipole optical antenna (depicted by red arcs). (b) Typical BLS spectrum schematically plotted in a semi-log \textit{y}-axis. }
\label{fig1}
\end{figure}

Our idea is schematically illustrated in Fig.~\ref{fig1}(a). The spectrum of light scattered from acoustic waves has a form of a triplet, consisting of the central Rayleigh peak and two Brillouin peaks shifted from the frequency of the incident light by the frequency of acoustic wave [Fig.~\ref{fig1}(b)] \cite{Fab68}. In conventional approaches to detecting BLS, the Brillouin shift and the intensity of the Brillouin peaks are much smaller than the frequency and intensity of the incident light, respectively, and sophisticated instrumentation \cite{Seb15, Pre12, Mon86} has to be used to determine them experimentally. To address this problem, we numerically simulate the BLS effect from acoustic waves and show that the intensity of the BLS signal can be increased by exploiting higher-order plasmon modes supported by the dipole optical antenna.

While there are many different optical antenna designs \cite{Nov11, Kra13}, we focus our attention on dipole antennae that consist of a single nanorod made of silver. We are interested not only in the response of the dipole antenna to the fundamental resonance, but also to higher-order plasmon modes. Our focus on higher-order modes is to ultimately monitor backscattered acoustic radiation from biological tissue and bodily fluids, but also to detect nonlinear acoustic responses from such media \cite{Hamilton}, which affords new sensing opportunities \cite{Ma15}. By using an ultrasmall optical antenna, rather than traditional mm-size, ultrasound transducers, we see new opportunities for IVPA imaging \cite{Wan10}.

The existence of higher-order plasmon modes of dipole optical antennae have been theoretically predicted and experimentally observed by several groups (see, e.g., Refs.~\onlinecite{Pay06, Khl07, Cha09, Wei10, Ree11, Abb12, Lop12, Ver15}). These modes have been shown to be analogous to the modes of a Fabry-Perot resonator. This analogy comes from the fact that these modes arise from multiple reflections between the two ends of the antenna \cite{Sch03, Wei10, Ver15}. Most significantly, lower radiative losses associated with these modes lead to a higher local field confinement as compared with the fundamental mode. In particular, this property is attractive for the detection of weak dielectric permittivity fluctuations of the liquid around the antenna caused by ultrasound. We will exploit this property in the remainder of the paper.

Finally, although we are interested in linear and nonlinear acoustic transduction, it is also important to stress that our antennae can demonstrate optical nonlinearities \cite{Boyd, Kra13}. By numerical simulations we demonstrate the possibility to optically excite the antenna by the near-infrared radiation but probe the acoustic waves with light at the frequency of the third harmonic (TH) of the optical excitation signal [Fig.~\ref{fig1}(a)]. This is achievable by designing the antenna such that the frequencies of one of its higher-order plasmon modes coincides with the frequency of the TH signal. Most significantly, a strong local field associated with the higher-order modes leads to a considerable enhancement of the TH signal in close vicinity to the antenna surface. This makes the antenna more sensitive to the fluctuations of the dielectric permittivity of the liquid that surrounds the antenna and transmits acoustic waves. Such nonlinear optical signal conversion can be advantageous in practical situations when the operation at shorter wavelengths is preferable due to specific absorption properties of biological tissues or liquids that are being insonified by ultrasound. Moreover, the third harmonic generation (THG) itself can be used as an independent bioimaging modality \cite{Zip03} that can complement the IVPA imaging.

The remainder of the paper is organised as follows. In subsection \ref{Model equations} we present a general theory of nonlinear acoustic wave propagation in bulk water, which is chosen as the model liquid surrounding the optical antenna. Based on this theory, we develop an acoustic finite-difference time-domain (a-FDTD) method, which allows calculating the spatial profiles of the scalar pressure field $p(\bm{r},t)$ and the vector velocity field $\bm{v}(\bm{r},t)$ around the antenna. These profiles allow us to understand the properties of the optical antenna to scatter acoustic waves. In subsection \ref{Fluctuations of the dielectric permittivity of liquid caused by acoustic waves}, from the simulated profiles of the acoustic fields we extract the acoustically-driven dynamics of the fluctuations of the dielectric permittivity of water, which is then used in our in-house optical finite-difference time-domain (o-FDTD) method modified to simulate the BLS spectra by taking into account the extremely slow dynamics of acoustics waves as compared with light. In subsection \ref{Plasmon-enhanced Brillouin light scattering}, we demonstrate the enhancement of the BLS signal due to the higher-order plasmon modes of the antenna. Finally, in Section \ref{Design of the optical antenna}, we discuss a potential practical scheme of the excitation of the higher-order modes of the antenna by means of nonlinear optical effects. We also present numerical details of the a-FDTD and o-FDTD methods in Appendix A and Appendix B, respectively.

\section{Results and discussion} \label{Results and discussion}

We explore the plasmon-assisted enhancement of the intensity of peaks arising in the spectral response of the optical antenna due to the BLS from acoustic waves, which propagate in bulk liquid that surrounds the antenna. To model the optical response of the antenna we employ the finite-difference time-domain (FDTD) method that is one of the standard numerical approaches employed in electrodynamics and photonics \cite{sullivan}. 

Finite-difference algorithms are also often used to solve acoustic wave propagation and scattering \cite{sullivan}. Conceptually, the acoustic FDTD (a-FDTD) is similar to optical FDTD (o-FDTD). This is because both approaches are based on first-order differential equations where the temporal derivative of one field is related to the spatial derivative of another field.  Thus, it is theoretically possible to merge the two FDTD algorithms to self-consistently solve the problem of BLS from acoustic waves in the presence of the optical antenna.

In practice a rigorous self-consistent numerical solution that combines o-FDTD and a-FDTD is very challenging because the acoustic frequency (e.g., $10$ MHz) and the speed of sound (e.g, $1500$ m/s) are widely disparate from those of light ($150-600$ THz and $3 \times 10^{8}$ m/s). Hence, the numerical FDTD model effectively has two very different time scales. The first time scale is at the rapidly varying electromagnetic (light) frequency. The second time scale is the slowly varying  acoustic frequency. As a result, the electromagnetic solution reaches a steady state solution before the liquid state has changed substantially due to the propagation of the acoustic wave.

Inspired by the theoretical approach presented in Ref.~\onlinecite{Mou66}, we circumvent the problem of different time scales by considering hypersonic acoustic waves that have GHz frequencies \cite{Fab68}, instead of ultrasonic wave having several orders of magnitude lower frequencies. In many liquids, such waves can be excited by a laser pulse \cite{Sto98} or piezoelectric film transducers \cite{Zho11}. Importantly, in liquids the general properties of hypersonic waves are similar to those of ultrasonic waves. Hence, the propagation of hypersonic waves can be studied by using conventional acoustic approaches \cite{Fab68}. In the framework of our numerical model, the consideration of hypersonic waves instead of ultrasonic ones allows us to reduce, to some extent, the mismatch between the time scales of the o-FDTD and a-FDTD models. This eventually makes a full-vectorial three-dimensional $3$D numerical model solvable when using high-performance computers. However, a more systematic analysis carried out in this paper has required us to additionally reduce the dimensionality of the acoustic model to two-dimensions ($2$D).

For our purposes, we assume that our antenna is made of silver and it is surrounded by bulk water. This assumption is done without loss of generality because silver is often used as the model material of optical antennae, and water is present in many practical situations including the propagation of ultrasound in biological objects. Moreover, silver is used as the constituent material of nanoparticles employed as contrast agents for IVPA imaging \cite{Hom12}. 

The propagation of acoustic waves in water (as well as in blood, fat, and other biological tissues and liquids) is governed by the equations of nonlinear acoustics derived from the full system of governing equations of fluid dynamics \cite{Hamilton}. Indeed, differences between water and other fluids and materials is likely to be the ultimate sensing target of a final device and could, in principle, be treated with the approaches we outline below.

The speed of the longitudinal acoustic wave $c_{\rm{0}}$ in water (silver) is $1500$ m/s ($3650$ m/s) and the density of water (silver) is $1000$ kg/m$^3$ ($10500$ kg/m$^3$). These parameters are taken into account by the model equations of nonlinear acoustics in liquids.

However, silver also support transverse acoustic waves, the speed of which is smaller than that of longitudinal waves. The speed of transverse waves is a material parameter taken into account in the equations of elastodynamics, which have to be solved if acoustic vibrations of the optical antenna are of interest. It is noteworthy that in optical antennae the frequency of these vibrations is $\sim 10-20$ GHz or even higher, which depends on the antenna dimensions \cite{OBr14}. Hereafter, we neglect such vibrations, and thus we do not consider the elasticity equations in our model, which assumes that the water acoustic waves interact with the silver antenna as with a rigid body. This approximation is warranted because the acoustic impedance of silver is much higher than that of water \cite{Beranek}.  Another reason to neglect the acoustic vibrations of the antenna is the fact that we consider GHz-range hypersonic waves instead of ultrasonic waves (see above). We project the results obtained for the hypersonic waves to the case of acoustic waves having lower frequencies, at which the acoustic vibrations of the ultrasmall antenna are negligible. This latter approximation is also warranted because the dimensions of the antenna are small as compared with the wavelength of acoustic vibrations in the kHz and MHz spectral ranges. Indeed, acoustic vibrations at these frequencies are possible in optomechanical micro-cavities whose characteristic dimensions reach $10-50$ $\mu$m \cite{vahala}, i.e. $\sim 30-150$ times larger than an optical antenna.

\subsection{Model equations of nonlinear acoustics in liquids} \label{Model equations}

We begin with the continuity equation, which links the vector velocity vector $\bm{v}$ to the instantaneous density of liquid $\rho$ as
\begin{align}
\frac{\partial \rho}{\partial t} + ( \bm{v} \cdot \bm{\nabla}) \rho + \rho \bm{\nabla} \cdot \bm{v} = 0
\label{eq:one},
\end{align}

\noindent where $t$ is the time.

The second equation is the momentum equation:
\begin{align}
\begin{aligned}
\rho \left( \frac{\partial \bm{v}}{\partial t} + ( \bm{v} \cdot \bm{\nabla}) \bm{v} \right) {}& = \\
 - \bm{\nabla}P  + \eta \nabla^{2} \bm{v} {}& + (\zeta + \frac{\eta}{3}) \bm{\nabla} (\bm{\nabla} \cdot \bm{v}),
\end{aligned}
\label{eq:two}
\end{align}
\noindent where $P$ is the thermodynamic pressure, $\eta$ the shear viscosity, and $\zeta$ the bulk viscosity.

The third, the thermodynamic equation of state, relates three physical quantities describing the thermodynamic behaviour of the liquid. These quantities can be $P$, $\rho$, and the temperature $T$. Alternatively, they can be $P$, $\rho$, and the specific entropy $s$. Normally, one applies a Taylor expansion about the equilibrium state, which to second order is 
\begin{align}
\begin{aligned}
p = c_{\rm{0}}^{2} \rho' + \frac{c_{\rm{0}}^{2}}{\rho_{\rm{0}}} \frac{B}{2A} \rho'^{2} +\left(\frac{\partial P}{\partial s} \right)_{\rm{\rho,0}}s'
\end{aligned}
\label{eq:three},
\end{align}
\noindent where $p$, $\rho'$, and $s'$ are the dynamic pressure, density, and entropy, respectively. These quantities describe small disturbances relative to the uniform state of rest. The parameters $B/A$ and $c_{\rm{0}}$ are the nonlinear parameter of the medium and the `small signal' speed of sound, respectively.

The fourth equation is the entropy equation
\begin{align}
\begin{aligned}
\rho T \left( \frac{\partial s}{\partial t} + ( \bm{v} \cdot \bm{\nabla}) s \right) = \kappa \nabla^{2} T + \zeta(\bm{\nabla} \cdot \bm{v})^{2} +\\
\frac{\eta}{2} \left(\frac{\partial v_{\rm{i}}}{\partial x_{\rm{j}}} + \frac{\partial v_{\rm{j}}}{\partial x_{\rm{i}}} -\frac{2}{3} \delta_{\rm{i,j}} \frac{\partial v_{\rm{k}}}{\partial x_{\rm{k}}} \right)^{2},
\end{aligned}
\label{eq:four}
\end{align}
\noindent where $\kappa$ is the thermal conductivity and $\delta_{\rm{i,j}}$ is the delta function. The last term of the right hand side of Eq.~\ref{eq:four} is written in  Cartesian tensor notation. 

By keeping the first and second order terms in Eqs.~\ref{eq:one}-\ref{eq:four}, we obtain the following nonlinear equations. The continuity equation is
\begin{align}
\begin{aligned}
\frac{\partial \rho'}{\partial t} + \rho_{\rm{0}} \bm{\nabla} \cdot \bm{v} = \frac{1}{\rho_{\rm{0}} c_{\rm{0}}^{4}} \frac{\partial p^{2}}{\partial t} + \frac{1}{c_{\rm{0}}^{2}} \frac{\partial \mathcal{L}}{\partial t}
\end{aligned}
\label{eq:five},
\end{align}
\noindent and the momentum equation is reexpressed as
\begin{align}
\begin{aligned}
\rho_{\rm{0}} \frac{\partial \bm{v}}{\partial t} + \bm{\nabla} p = - \frac{1}{\rho_{\rm{0}} c_{\rm{0}}^{2}} (\zeta + \frac{4\eta}{3}) 
\bm{\nabla} \frac{\partial p}{\partial t} - \bm{\nabla} \mathcal{L}
\end{aligned}
\label{eq:six},
\end{align}
\noindent where $\mathcal{L} = \frac{1}{2} \rho_{\rm{0}} v^{2} - \frac{p^{2}}{\rho_{\rm{0}} c_{\rm{0}}^{2}}$ is the second-order Lagrangian density. The entropy equation can be simplified to
\begin{align}
\begin{aligned}
\rho_{\rm{0}} T_{\rm{0}} \frac{\partial s'}{\partial t} = \kappa \nabla^{2} T
\end{aligned}
\label{eq:seven},
\end{align}
\noindent where $T_{\rm{0}}$ the temperature of the medium at rest. The substitution of Eq.~\ref{eq:seven} to Eq.~\ref{eq:three} and the use of thermodynamic relationships, one can write the equation of state as
\begin{align}
\begin{aligned}
\rho' = \frac{p}{c_{\rm{0}}^{2}} - \frac{1}{\rho_{\rm{0}} c_{\rm{0}}^{4}} \frac{B}{2A} p^{2} - \frac{\kappa}{\rho_{\rm{0}} c_{\rm{0}}^{4}} \left(\frac{1}{c_{\rm{v}}} - \frac{1}{c_{\rm{p}}} \right) \frac{\partial p}{\partial t}
\end{aligned}
\label{eq:eigth},
\end{align}
\noindent where $c_{\rm{v}}$ and $c_{\rm{p}}$ are the specific heat capacities at constant volume and pressure, respectively. We note that for a variety of liquids the constants used in Eqs.~\ref{eq:one}-\ref{eq:eigth} have been tabulated in the literature \cite{tables}.

It is also important to derive the linearised acoustic equations. These equations serve as the departure point for the derivation of the discrete finite-difference analogues of nonlinear equations Eqs.~\ref{eq:one}-\ref{eq:eigth}. In many cases it is instructive to start the analysis with the solution of the linear equations, and then to proceed with the solution of the nonlinear equations, as the linearised equations are easier to discretise and solve using the basic finite-difference formulations than their nonlinear counterparts \cite{Wan96, sullivan}. Additionally, the linear equations are also important because at the boundaries of the computational domain it is technically impossible to impose absorbing boundary conditions when using the nonlinear equations \cite{sullivan}. The absorbing boundary conditions is a standard approach that is used in the framework of the FDTD method to simulate the propagation of waves either in free space or in a medium in which the waves are fully attenuated before they can experience reflection from boundaries.

Keeping only the term of the first order, the linearised continuity equation Eq.~\ref{eq:one} becomes
\begin{align}
\begin{aligned}
\frac{\partial \rho'}{\partial t} + \bm{\nabla} \cdot (\rho_{\rm{0}} \bm{v}) = 0
\end{aligned}
\label{eq:A1},
\end{align}
\noindent where $\rho_{\rm{0}}$ is the density of the undisturbed medium. Similarly, the momentum equation Eq.~\ref{eq:two} becomes 
\begin{align}
\begin{aligned}
\rho_{\rm{0}} \frac{\partial \bm{v}}{\partial t} = - \bm{\nabla} p
\end{aligned}
\label{eq:A2}.
\end{align}
The linearised equation of state Eq.~\ref{eq:three} becomes
\begin{align}
p = c_{\rm{0}}^{2} \rho'
\label{eq:A3}.
\end{align}

Our discrete numerical model is based on the standard finite-difference stencils employed in the FDTD method \cite{sullivan}. The finite-difference analogues of both linear and nonlinear model equations are given in Appendix A.

\subsection{Fluctuations of the dielectric permittivity of liquid caused by acoustic waves} \label{Fluctuations of the dielectric permittivity of liquid caused by acoustic waves}

The simulation of  BLS from an acoustic wave requires one to define the time dependences of the fluctuations of the dielectric permittivity cased by the acoustic wave propagating in the liquid. This can be done by calculating the profiles of the acoustic scalar pressure field \textit{p} and the vector velocity field \textbf{v}. To start with, we calculate these profile for the linear case, i.e. we neglect the nonlinear acoustic effects. Once the behaviour of the dielectric permittivity in the linear case has been understood, we will take into account the nonlinear effects.

So far, we have not discussed the particular design of the optical antenna. We assume that the antenna consists of a silver nanorod, which has a square-cross section $w=30$~nm and length $340$~nm. This choice of  parameters is due to optimisation using o-FDTD simulations, and is discussed in more detail in Section \ref{Design of the optical antenna}. 

In the a-FDTD model, the antenna is insonified by a plain quasi-monochromatic acoustic pressure wave propagating in bulk water. The peak amplitude of this wave is $5$~MPa, which is consistent with the pressure level used in nonlinear intravascular contrast imaging \cite{Goe06}. In the time-domain, the profile of this wave is given by a very long Gaussian pulse modulated by a sinusoidal wave. The pulse length was $5000 \Delta_{\rm{t}}$ where $\Delta_{\rm{t}}$ is the time step in the a-FDTD algorithm.

\begin{figure}[tb]
\centering\includegraphics[width=6.5cm]{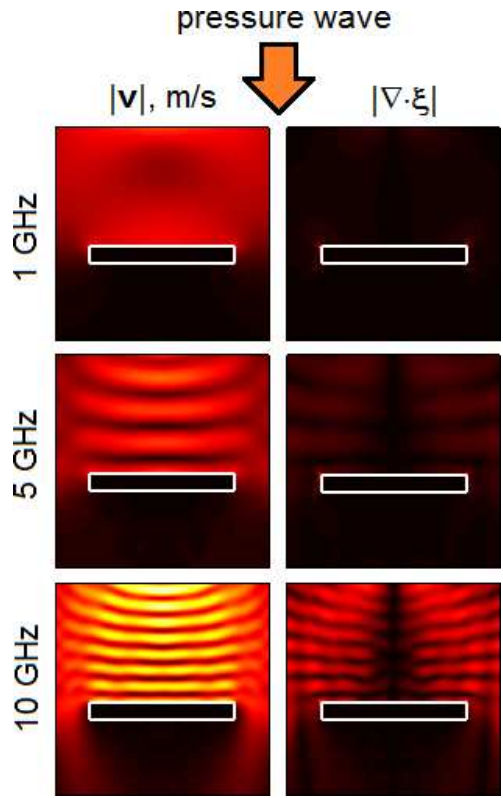}
\caption{Simulated spatial profiles of the velocity $|\bm{v}|$ (left column) and the divergence of the displacement field $|\bm{\xi}|$ for the acoustic frequency $1$~GHz, $5$~GHz, and $10$~GHz. The white rectangle denotes the contours of the antenna and also serves as the scalebar with dimensions $340$~$\times$~$30$~nm. The black (yellow) colour denotes the zero (maximum) of the profile intensity. The direction of the incident monochromatic pressure wave is indicated by the arrow. Note that the velocity $|\bm{v}|$ in front of (behind) the antenna increases (decreases), which also changes the profile of $\nabla \cdot \xi$. This is because of the pressure doubling effect \cite{Beranek}.}
\label{fig2}
\end{figure}

The left column of Fig.~\ref{fig2} shows the simulated profiles of $|\bm{v}|$ in close proximity to the antenna for the acoustic frequencies $f_{\rm{a}} = 1$, $5$, and $10$ GHz. These profiles were obtained by Fourier-transforming the dependencies $\bm{v}(\bm{r},t)$ produced by the a-FDTD method. The contours of the antenna are outlined by the white $340$ nm $\times$ $30$ nm rectangle, which also serves as the scalebar.  We also calculate the displacement field $\xi$, which is the displacement of a particle from its equilibrium position due to the acoustic wave,  at the same frequencies as above by  $\xi = \int_{\rm{t}}\bm{v}(\bm{r},t) \rm{d}t$ \cite{Beranek}. The calculation of $\xi$ by means of the a-FDTD method is relatively simple when the incident signal is quasi-monochromatic. 

As a next step, we calculate the divergence of the displacement field $\nabla \cdot \xi$, as shown in the right column of Fig.~\ref{fig2}. The amplitude of dynamic fluctuations of the dielectric permittivity of water $\delta \epsilon$ caused by the propagating acoustic wave is directly proportional to $\nabla \cdot \xi$, as shown in Ref.~\onlinecite{Law01}. This is because in the case of the plain longitudinal acoustic waves, the liquid is alternately compressed and stretched, which gives rise to alternating density changes. The knowledge of the changes in the dielectric permittivity is required to simulate the BLS from acoustic waves by means of the o-FDTD method. We estimate that in our model the amplitude of $\delta \epsilon$ is $\sim 10^{-4}$.

One can see that the spatial profile of $\nabla \cdot \xi$ is quasi-uniform at the acoustic frequency $f_{\rm{a}} = 1$ GHz. This is because the wavelength of the acoustic wave in water is $\sim 5$ times larger than the dimensions of the antenna. As a result, the acoustic wave is weakly scattered by the antenna. Consequently, the profile of the velocity $\bm{v}$ is quasi-uniform, which implies that in close proximity to the antenna $\nabla \cdot \xi \approx 0$. Hence, the profile of $\delta \epsilon$ is quasi-uniform, too. The same result holds for lower frequencies (not shown). 

However, at higher frequencies the profile of $\nabla \cdot \xi$ becomes less uniform. This can be seen at $\sim 5$ GHz, i.e. when the acoustic wavelength $300$ nm is close to the length of antenna. Moreover, at frequencies $>5$ GHz the profile is no longer uniform (Fig.~\ref{fig2}, right column). Consequently, if the frequency is $<5$ GHz, which is the case of our study, we will abstract from the fact that we simulate the propagation of the GHz-range hypersonic waves. Instead, we will consider the normalised acoustic frequency $f_{\rm{a}}/f_{\rm{a,0}}$, where $f_{\rm{a,0}}$ is the frequency of the fundamental harmonic of the incident wave. This approach allows us to project our results to the case of ultrasonic waves.

We now consider nonlinear acoustic effects. To do so, in our a-FDTD software we turn on the acoustic nonlinearity by setting $\beta = 1 + \frac{B}{2A} = 3.5$, which corresponds to the nonlinear acoustic parameter of water \cite{Hamilton}. As shown in Appendix A, the nonlinearity is present in the $1$D model only. The input quasi-monochomatic pressure wave has the frequency $f_{\rm{a,0}}$ and amplitude $5$~MPa. The distortion due to the nonlinear propagation signal is detected at the output of the $1$D model and used as the excitation source in the same $2$D model employed to produce the results in Fig.~\ref{fig2}. In general, the distorted signal carries energy at the fundamental frequency $f_{\rm{a,0}}$ as well as at all harmonics generated due to the nonlinear interaction. It is noteworthy that the amplitude of the generated harmonics depends on the wave propagation distance in the liquid. In our model, this distance equals approximately one quarter of $x_{\rm{p}} = \frac{\lambda_{\rm{a,0}}}{2 \pi \beta M}$, where $\lambda_{\rm{a,0}}$ is the wavelength of the fundamental wave and $M$ the acoustic Mach number. The value of $x_{\rm{p}}$ corresponds to the distance at which the amplitude of the second nonlinear generated harmonics formally reaches $1/2$ of the amplitude of the fundamental wave \cite{Rudenko_Russian}.

\begin{figure}[tb]
\centering\includegraphics[width=8.5cm]{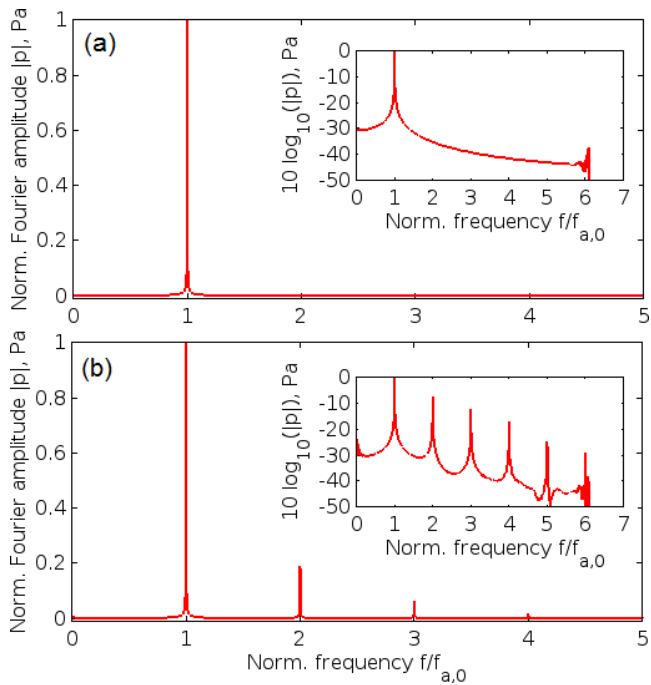}
\caption{ Normalised acoustic energy spectrum of the incident quasi-monochormatic pressure wave detected (a) in the input port of the $1$D model and (b) in output port after the propagation in water.  The frequency of the incident wave is $f_{\rm{a,0}}$ and its amplitude is $5$ MPa. The insets show the same spectra plotted in the decibel scale. Note that in the nonlinear spectrum in total five new frequency harmonics can be seen above the $-40$ dB level. Furthermore, one can distinguish the dc component (zero frequency) generated due to the nonlinear acoustic effect.}
\label{fig3}
\end{figure}

The simulated spectrum of the linear acoustic wave is shown in Fig.~\ref{fig3}(a). It was obtained by Fourier-transforming the calculated time-domain dependencies of the pressure $p(t)$ detected in the input port of the $1$D a-FDTD model. From Fig.~\ref{fig3}(a) and its inset, which shows the same spectrum plotted in the decibel scale, one can see only one peak at the frequency $f_{\rm{a,0}}$ confirming the monochromatic nature of this wave.

However, when the acoustic nonlinearity is taken into account, in Fig.~\ref{fig3}(b) one can see new nonlinear generated waves appear in the spectrum at the frequencies $2f_{\rm{a,0}}$ (quadratic nonlinear effect), $3f_{\rm{a,0}}$ (cubic effect), etc. This spectrum was obtained by Fourier-transforming the values of $p(t)$ detected in the output port of the $1$D model. In addition to the harmonic waves, the nonlinear interaction leads to the creation of a dc component (i.e. zero frequency), which can be seen upon a close inspection of the inset in Fig.~\ref{fig3}(b). We note that the relative amplitudes of the nonlinear generated waves are in good agreement with the predictions of analytical formulae presented in Ref.~\onlinecite{Rudenko_Russian}.

In the nonlinear case, we also extract the time-domain dependencies of the dielectric permittivity fluctuations $\delta \epsilon(t)$ (not shown). In contrast to the linear case, these dependencies contain the contributions from the density changes caused by both the fundamental wave and its harmonics. The contributions of the harmonic waves are small as compared with those of the fundamental wave because their relative magnitudes follow the intensities of the spectral components shown in Fig.~\ref{fig3}(b). We use $\delta \epsilon(t)$ in the simulations of light scattering from nonlinear acoustic waves. The results of these simulations are presented in subsection \ref{Plasmon-enhanced Brillouin light scattering}.     

\subsection{Plasmon-enhanced Brillouin light scattering} \label{Plasmon-enhanced Brillouin light scattering}

We demonstrate a plasmon-enhanced light-acoustic wave interaction in the presence of the antenna. The a-FDTD simulations conducted in the previous Section \ref{Fluctuations of the dielectric permittivity of liquid caused by acoustic waves} have produced the time dependences of the dielectric permittivity fluctuations at each discrete point inside the computational domain. We correlate these points with the discrete points of the computational domain in the o-FDTD method, as explained in Appendix B.

We use the theory from Refs.~\onlinecite{Khl07, Nov11, Lop12} to estimate the frequency of the fundamental (dipole) plasmonic mode of the antenna: $\sim 2100$ nm. The theory also predicts that this antenna should support two higher-order plasmonic modes at $\sim 700$ and $\sim 500$ nm, respectively.

Hereafter, we also assume that the acoustic waves in water are probed at the frequency of the first higher-order plasmonic mode at $\sim 700$ nm. The reason for this is twofold. As will be shown in Section \ref{Design of the optical antenna}, the electric field profiles of the higher-order plasmonic modes exhibit Fabry-Perot-like behaviour and are tightly confined to the surface of the antenna. These modes bounce back and forth as they experience multiple reflections from the ends of the antenna. This behaviour is favourable for the detection of acoustic waves because the multiple passing of light along the antenna surface effectively increases the interaction time of light with the acoustic wave. Furthermore, the tight confinement of the electric field gives rise to an increased sensitivity of the plasmon resonance to any change in the dielectric permittivity occurring in close proximity to the surface of the antenna \cite{Nus07}.

It is important to mention that for numerical tractability we have to make several simplifications, which do not interfere with our analysis but drastically decrease simulation time and simplify the analysis of the plasmon-enhanced BLS effect. 

Firstly, we limit ourselves to the $2$D o-FDTD model. Apart from the fact that the a-FDTD model is also $2$D, this simplification is warranted because $2$D o-FDTD simulations produce qualitatively the same linear spectra as those produced using a $3$D model (see, e.g., Ref.~\onlinecite{Mak11_opt_commun}).

Secondly, we assume that the optical signal exciting the antenna does not generate additional acoustic waves in water. This is because an efficient generation of acoustic waves in water is possible only when one fulfils certain conditions \cite{Hom12, Sto98}. However, those conditions are not met in our model because their fulfilment is out of the scope of this work. Furthermore, we also neglect the impact of the radiation pressure, electrostriction, thermal expansion, and other effects that potentially can give rise to acoustic vibrations of the antenna due to the interaction with the optical radiation.

\begin{figure}[tb]
\centering\includegraphics[width=8.5cm]{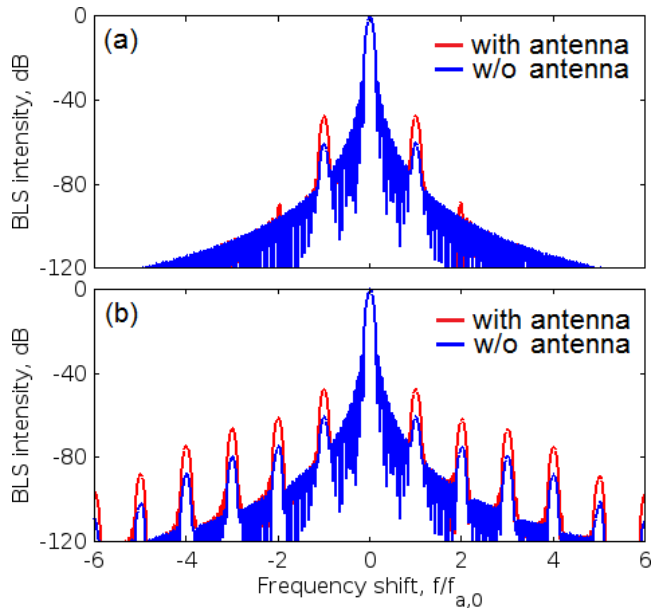}
\caption{Simulated plasmon-enhanced intensity of light scattered from (a) the monochromatic sinusoidal acoustic wave and (b) nonlinear generated acoustic waves. Note a $\sim 35$-fold enhancement due to the presence of the nanorod in Panel (a).}
\label{fig4}
\end{figure}

To start with, we simulate the effect of BLS from a quasi-monochromatic sinusoidal acoustic wave of frequency $f_{\rm{a,0}}$ propagating in the linear medium. In the a-FDTD software, we turn off the nonlinearity by setting $\beta = 0$. We run the a-FDTD simulation in parallel with the o-FDTD simulation, where the latter receives time-dependent information about the dielectric permittivity fluctuations of water caused by the propagating acoustic wave. Two batches of simulations are conducted: one with the optical antenna in bulk water and another one  in bulk water without the antenna. The results are presented in Fig.~\ref{fig4}(a). In the case without the antenna (blue curve) one can see a typical BLS spectrum with the the central Rayleigh peak and two Brillouin peaks shifted from the frequency of the incident light by the frequency of acoustic wave $f_{\rm{a,0}}$.

Importantly, in the presence of the antenna one can see that the intensity of Brillouin peaks is enhanced (red curve). This $\sim35$-fold increase with respect to the BLS intensity without the antenna is attributed to the plasmon resonance of the antenna. To verify this, we conducted an extra simulation (results not shown here) in which the frequency of the optical signal that excites the antenna was off-resonance. As expected, that simulation produced virtually the same result as without the antenna.

We also note the existence of two additional Brillouin peaks shifted from the frequency of the incident light by $2f_{\rm{a,0}}$ [Fig.~\ref{fig4}(a)]. As the acoustic nonlinearities were neglected in the simulation, the appearance of these very weak peaks is attributed to the cascaded scattering of the optical signals at the frequency $\pm f_{\rm{a,0}}$ from the acoustic wave. In this case, the optical waves at $\pm f_{\rm{a,0}}$ act as the incident light but the signals at $\pm 2f_{\rm{a,0}}$ are the Brillouin-shifted waves, which are very weak and visible only in the presence of the antenna.

Finally, we simulate the BLS from nonlinear acoustic waves in the presence of the antenna. We repeat the same self-consistent acoustic and optical simulations as above, but this time we take into account the nonlinear acoustic properties of water. The result of this simulation is presented in Fig.~\ref{fig4}(b). In contrast to the BLS effect from the linear wave [Fig.~\ref{fig4}(a), red curve], in Fig.~\ref{fig4}(b) one can see multiple Brillouin peaks corresponding to the scattering from nonlinear generated harmonics of the incident acoustic wave. Importantly, the enhancement of the BLS intensity by the antenna occurs only at the frequencies of the Brillouin peaks and it does not affect a broad Rayleigh-wing background. This can be seen by inspecting the difference between the red and blue curves in Fig.~\ref{fig4}(b) that correspond to the cases with and without the antenna, respectively. This selective enhancement will help to resolve the fine structure in BLS spectra.

It is noteworthy that the optical spectrum in Fig.~\ref{fig4}(b) resembles a Brillouin frequency comb similar to that in Ref.~\onlinecite{cudos}. However, the physical origin of this frequency comb is different. This comb is generated due to the BLS from the nonlinear acoustic waves as well as the subsequent plasmon-assisted enhancement of the BLS peaks. We verified that the intensity of the individual optical spectral components in Fig.~\ref{fig4}(b) corresponds to the intensity of the harmonics in the acoustic spectrum in Fig.~\ref{fig3}(b). We note that the spectral composition of the frequency comb in Fig.~\ref{fig4}(b) can be controlled by tailoring the nonlinear acoustic interaction in water. This functionality can be exploited in recently proposed Brillouin cavity opto-mechanic systems combined with microfluidic devices \cite{Bah13}. 

\section{Design of the optical antenna} \label{Design of the optical antenna}

We discuss the possible design of the optical antenna and one of the possible practical schemes for its optical excitation. In the discussion above, we have already mentioned the existence of higher-order plasmon modes supported by the antenna. We also assumed that in our simulations of the plasmon-enhanced BLS effect, the acoustic waves were probed by light whose frequency corresponds to the resonance frequency of the excited higher-order mode. The frequency of this mode was estimated using simple effective wavelength theory.

Recall that the higher-order modes of the antenna are analogous to Fabry-Perot modes because they arise from multiple reflections between the two ends of the silver nanorod. It is noteworthy that different Fabry-Perot structures and other types of optical resonators are often used in many practical situations involving the detection of acoustic waves with light (see, e.g., Ref.~\onlinecite{Li14} and references therein). This is because the presence of the resonator effectively increases the interaction time of light with the acoustic waves. We also note that the resonant excitation of plasmons has been used to additionally improve the resolution and sensitivity of the BLS spectroscopy and relevant techniques (see, e.g., Refs.~\onlinecite{Ute08, Nus07}). Consequently, the higher-order modes effectively combine both the resonant excitation of plasmons with the Fabry-Perot-like behaviour.

We also remind that the far-field optical radiation properties of the higher-order plasmonic modes of the antenna are poor as compared with those of the dipole modes. By virtue of the principle of reciprocity \cite{Nov11}, this implies that the efficiency of the excitation of the higher-order modes from the far-field region is also relatively low. Consequently, these modes typically have to be excited by near-field optical radiation, for example, by using a point-like quantum nano-emitter such as a quantum dot \cite{Nov11}.   

However, there are alternatives to the near-field excitation scheme involving nano-emitters. We demonstrate that the higher-order modes can be excited with the far-field radiation via nonlinear third harmonic generation. Previously, Abb \textit{et. al.} \cite{Abb12} demonstrated a nonlinear control of the higher-order modes in asymmetric nanorod antenna with a gap. Here, we show that the light intensity enhancement in the near-field region associated with the excitation of these modes leads to a considerable increase of the THG effect as compared with the case without the antenna. This distinguishes our approach from that used in many previous works that exploit dipole plasmonic modes to enhance nonlinear optical effects at the nanoscale (see, e.g., Refs.~\onlinecite{Pal08, Thy12, Aou14}).

\begin{figure}[tb]
\centering\includegraphics[width=8.5cm]{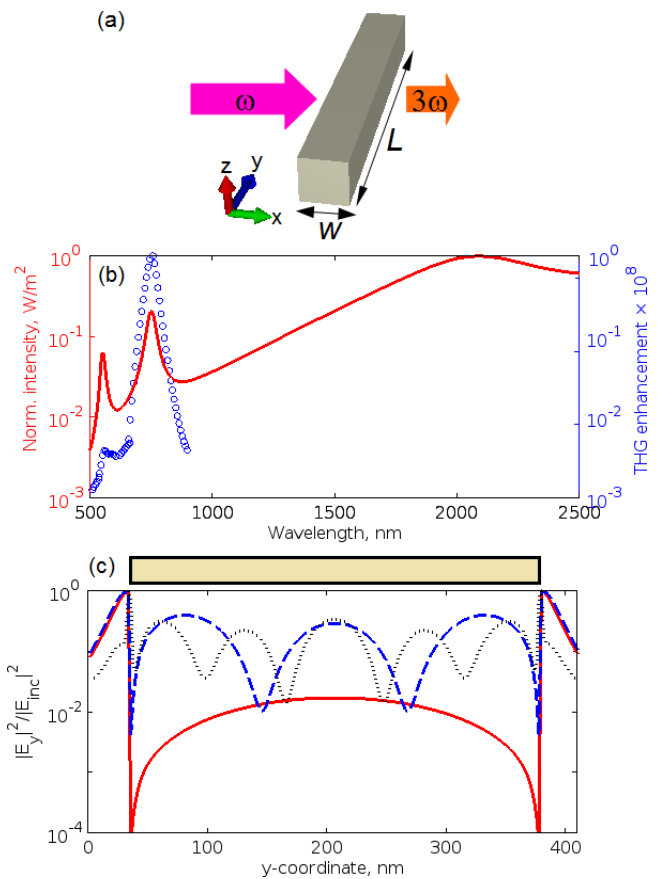}
\caption{(a) Illustration of the numerical simulation configuration for the investigation of the THG in the antenna. A plane electromagnetic wave is normally incident from the left on the antenna. The electric field vector of the plane wave is parallel to the long axis of the antenna. The generated TH signal, as well as the signal and the fundamental frequency, are probed from the right in a point located $50$ nm off the middle point of the nanorod. (b) Left \textit{y}-axis: Normalised to unity linear optical intensity spectrum of the antenna. Right \textit{y}-axis: THG enhancement spectrum. (c) Normalised simulated electric field intensity through the centre of the antenna cross-section. The rectangle above Panel (c) schematically indicates the length of the antenna.}
\label{fig5}
\end{figure}

We continue considering the silver antenna with a square cross-section $w = 30$ nm and length $L=340$ nm. However, since optical simulations in the absence of acoustic waves are significantly faster, we conduct rigorous $3$D numerical simulations of the linear and nonlinear response of the antenna. The configuration of our model is shown in Fig.~\ref{fig5}(a). The antenna is embedded into a uniform loss-less dielectric medium with an effective dielectric constant $\epsilon_{\rm{eff}}=1.75$. Whereas this value is close to the optical dielectric permittivity of water, it also accounts for the presence of a dielectric substrate of the antenna or an optical fibre use to guide light towards the antenna. These components can be present in a realistic device.     

Figure~\ref{fig5}(b) (red curve) shows the simulated linear spectrum of the antenna. One can see a broad dipole mode peak at $\sim 2000$ nm and two narrow higher-order mode peaks at $750$ nm and $550$ nm. Remarkably, the spectral position of the observed peaks is in good agreement with the approximate resonant wavelengths produced by a simple theoretical model based on the effective wavelength.

In Fig.~\ref{fig5}(c) we present the profiles of the normalised electric field intensity through the centre of the cross-section, i.e. along the $y$-axis in Fig.~\ref{fig5}(a). The red solid curve denotes the profile for the fundamental mode. The blue dashed (black dash-dotted) curve denotes the profile of the higher-order mode at $750$ nm ($550$ nm). The rectangle above Fig.~\ref{fig5}(c) schematically indicates the length of the antenna. One can see that the electric field profile of the higher-order plasmon modes resembles that for a Fabry-Perot resonator. 

Now we study the nonlinear optical THG effect arising due the optical antenna. We conduct simulations of the THG effect by using a $3$D nonlinear FDTD method \cite{Mak11}; for more details see Appendix B. We neglect the nonlinear susceptibility $\chi^{\rm{(3)}}$ of silver and for the surrounding effective dielectric medium we assume that $\chi^{\rm{(3)}}=1 \times 10^{\rm{-21}}$ m$^{\rm{2}}$/V$^{\rm{2}}$, which is close to the realistic value for water and other liquids \cite{Boyd}. Although the nonlinear susceptibility of silver can be higher than that of water, its neglect is justified by the fact that silver may be oxidised in the presence of water, which is expected to significantly reduce its contribution to the nonlinear response.

The antenna is excited in the $1200-2500$ nm spectral range with the amplitude of the electric field of $E_{\rm{y}} = 1.0 \times 10^{\rm{8}}$ V/m corresponding to the optical intensity of $\sim 2.3$ GW/cm$^{2}$. In our simulations, the intensity of the incident light is constant at all frequencies. This is a convenient theoretical approach that allows simplifying the analysis of results as compared with the case of the more realistic frequency-dependent intensity.

We study the THG enhancement defined as a ratio of the intensity of the generated TH signal in the presence of the antenna to that without it. We expect to observe a very large enhancement of several orders of magnitude because the intensity of the TH signal simulated in the absence of the antenna should be negligibly small due to a very short nonlinear optical interaction length. In our model this length equals $500$ nm. Indeed, as shown by open blue circles in Fig.~\ref{fig5}(b), the THG enhancement reaches a remarkably large value $\sim 10^8$ at the frequency of the first higher-order mode located at $750$ nm. This also corresponds to the conversion efficiency $\eta = 0.033\%$, which was calculated as the $I_{\rm{3\omega}}/I_{\rm{\omega}}$ being $I_{\rm{3\omega}}$ the intensity of the TH signal and $I_{\rm{\omega}}$ the intensity of the dipole mode of the optical antenna. We also note that even though the wavelength of the second higher-order peak ($550$ nm) is detuned from the wavelength of the TH signal, the THG enhancement spectrum at $550$ nm remains noticeable.

\section{Conclusions}

We have theoretically demonstrated the possibility to employ dipole optical antennae to enhance the scattering of light with nonlinear acoustic waves propagating in liquids. Although one can use the same optical antennae as those already employed in plasmon-enhanced photoacoustic imaging of biological tissues, we have showcased a different antenna design -- single $340$~nm-long silver nanorods that are long enough to support both fundamental and higher-order plasmon modes. We have shown that the higher-order plasmon modes of these nanorods lead to a strong interaction of light with acoustic waves. This effect originates from both the near-field enhancement due to the plasmon resonance and the Fabry-Perot-like behaviour of the antenna. Furthermore, we have shown the possibility to excite the optical antenna at one optical frequency but detect the acoustic waves at another optical frequency. This functionality is achievable by means of the nonlinear optical Kerr effect exhibited by water, which was considered as an ideal liquid that surrounds the antenna and also transmits the acoustic wave. We envision that this optical excitation scheme can find practical applications in biomedical imaging where different tissues and liquids have widely disparate optical absorption spectra. We also presented a hybrid acousto-optical finite-difference time-domain algorithm that allows one to investigate the interaction of light with nonlinear acoustic waves at the nanoscale.         

Furthermore, the optical antennae can be used to additionally increase the previously demonstrated benefits of the BLS spectroscopy technique from plasmon resonances in metallic nanostructures \cite{Ute08, Tem12, Mak15}. Ultrasound is also an important tool in food technology (see, e.g., Ref.~\onlinecite{Che11}), and optical antennae may be useful as a component of an all-optical system for noninvasive quality control of different processes involving ultrasonic waves. We also note that many high energy sonochemical reactions achievable with ultrasound \cite{sonochemistry} exploit the effect of cavitation, i.e. the formation, growth and collapse of bubbles in a liquid \cite{Hamilton}. Remarkably, the nonlinear acoustic coefficient of a bubbly liquid can be many orders of magnitude higher than that of the same liquid but without bubbles. Although the consideration of such strong acoustic nonlinearities goes far beyond the scope of this paper, we envision that the optical antenna can be used to facilitate the detection of highly nonlinear acoustic waves. 

\section{Acknowledgements}

This work was supported by Australian Research Council (ARC) through its Centre of Excellence for Nanoscale BioPhotonics (CE140100003). This research was undertaken on the NCI National Facility in Canberra, Australia, which is supported by the Australian Commonwealth Government. The authors would like to thank Andrey Sukhorukov for useful discussions.

\appendix

\section{The acoustic FDTD model}

The derivation of the discrete nonlinear acoustic equations is difficult as compared with the case of the linearised equations. Consequently, it is instructive to derive the discrete analogues of the linearised equations first, and then to use the resulting equations as the departure point for the derivation of the nonlinear equation. Using the FDTD notation proposed by Yee \cite{sullivan}, in the two-dimensional ($2$D) $x-y$ space our linear FDTD equations for the vector velocity $\bm{v}$ are written as
\begin{align}
\begin{aligned}
v_{\rm{x}}^{n+\frac{1}{2}}(i,j) &= v_{\rm{x}}^{n-\frac{1}{2}}(i,j) + \frac{K_{\rm{v}}}{\Delta_{\rm{y}}} \left[p^{n}(i,j) - p^{n}(i,j-1) \right] \\
v_{\rm{y}}^{n+\frac{1}{2}}(i,j) &= v_{\rm{y}}^{n-\frac{1}{2}}(i,j) -  \frac{K_{\rm{v}}}{\Delta_{\rm{x}}} \left[p^{n}(i,j) - p^{n}(i-1,j) \right]
\end{aligned}
\label{eq:A4},
\end{align}
\noindent where the index $n$ denotes the current iteration number. For the scalar pressure field $p$ we obtain 

\begin{align}
\begin{aligned}
p^{\rm{n+1}}(i,j) = &p^{\rm{n}}(i,j) \\
&+ \frac{K_{\rm{p}}}{\Delta_{\rm{y}}} \left[v_{\rm{x}}^{n+\frac{1}{2}}(i,j+1) - v_{\rm{x}}^{n+\frac{1}{2}}(i,j) \right] \\
&- \frac{K_{\rm{p}}}{\Delta_{\rm{x}}} \left[ v_{\rm{y}}^{n+\frac{1}{2}}(i+1,j) - v_{\rm{y}}^{n+\frac{1}{2}}(i,j) \right]
\end{aligned}
\label{eq:A5}.
\end{align}

Eqs.~\ref{eq:A4}-\ref{eq:A5} include the co-ordinate dependent coefficients $K_{\rm{v}}=\Delta_{\rm{t}}/\rho_{\rm{0}}$ and $K_{\rm{p}}=\Delta_{\rm{t}} \rho_{\rm{0}} c_{\rm{0}}^{2}$, where $\Delta_{\rm{t}} = \Delta_{\rm{x}}/(\sqrt{2} c_{\rm{0,max}})$ is the time step calculated assuming that $\Delta_{\rm{x}} = \Delta_{\rm{y}}$ and that $c_{\rm{0,max}}$ is the maximum speed of sound propagation in the liquid. Eqs.~\ref{eq:A4}-\ref{eq:A5} are solved iteratively in a fashion similar to electromagnetic FDTD solvers \cite{sullivan}. The information about how to model acoustically rigid bodies can be found in Ref.~\onlinecite{Wan96}.

It is noteworthy that in our simulations the nonlinear acoustic effects are taken into account as a one-dimensional ($1$D) model. This is because in many practical cases it suffices to solve a $1$D wave propagation problem in order to explain experimental results\cite{Vil13}. In addition, we do not expect that the presence of an ultra-small antenna to significantly perturb the strength of the nonlinear acoustic effects in water. Consequently, we use the $1$D nonlinear model to produce the frequencies and amplitudes of the harmonics of the original acoustic wave. As a next step of our simulations, these data are used in the linear $2$D model.

Following Ref.~\onlinecite{sullivan}, by using the first-order differencing in time and space, we derive the following finite-difference equations for the nonlinear $1$D case
\begin{align}
\begin{aligned}
v^{\rm{n+\frac{1}{2}}}(k)=v^{\rm{n-\frac{1}{2}}}(k)+g_{\rm{b}}\left[p(k)^{\rm{n}}-p^{\rm{n}}(k+1)\right] \\
- \frac{\zeta+\frac{4\eta}{3}}{\rho_{\rm{0}}^{2} c_{\rm{0}}^{2} \Delta_{\rm{z}}}\left[p^{\rm{n}}(k+1)-p^{\rm{n}}(k)-p^{\rm{n-2}}(k+1)+p^{\rm{n-2}}(k)\right]
\end{aligned}
\label{eq:A6},
\end{align}
\noindent where the constants $c_{\rm{0}}$, $\rho_{\rm{0}}$, $\zeta$, and $\eta$ are co-ordinate dependent, $g_{\rm{b}}=\Delta_{\rm{t}} \Delta_{\rm{z}}/\rho_{\rm{0}}$ and we also chose another co-ordinate index variable, $k$, and another step in space, $\Delta_{\rm{z}}$, in order to avoid the possible confusion with their counterparts in the $2$D model above. Importantly, as compared with the linear algorithm, the values of pressure $p$ at the two time steps -- $p^{\rm{n-2}}$ and $p^{\rm{n-1}}$ -- must be additionally retained in the computer memory. These extra values are required to update Eq.~\ref{eq:A6} as well as Eq.~\ref{eq:A7} which reads 

\begin{align}
\begin{aligned}
p^{\rm{n+1}}(k)= &-\frac{1}{\delta}\left\{\left[c_{\rm{0}}^{4} \Delta_{\rm{z}} p^{\rm{n-1}}(k) - 2 c_{\rm{0}}^{4} \Delta_{\rm{z}} p^{\rm{n}}(k)\right] \rho_{\rm{0}} C \right.\\
&+ c_{\rm{0}}^{4} \Delta_{\rm{t}}^2 \rho_{\rm{0}}^{2} \left[v^{\rm{n+\frac{1}{2}}}(k-1)-v^{\rm{n+\frac{1}{2}}}(k)\right] \\
&+ c_{\rm{0}}^{2} \Delta_{\rm{t}} \Delta_{\rm{z}} p(k) \rho_{\rm{0}} \\
&-\left.\beta \Delta_{\rm{t}} \Delta_{\rm{z}} p^{\rm{n}}(k) p^{\rm{n-1}}(k)+\beta \Delta_{\rm{t}} \Delta_{\rm{z}} p^{\rm{n}}(k)^{2}\right\}
\end{aligned}
\label{eq:A7},
\end{align}
  
\noindent where $ \delta = $ \tiny $\left[ c_{\rm{0}}^{4} \Delta_{\rm{z}} \rho_{\rm{0}} C - c_{\rm{0}}^{2} \Delta_{\rm{t}} \Delta_{\rm{z}} \rho_{\rm{0}}- \beta \Delta_{\rm{t}} \Delta_{\rm{z}} p^{\rm{n-1}}(k) + \beta \Delta_{\rm{t}} \Delta_{\rm{z}} p^{\rm{n}}(k) \right]$\normalsize, $\beta = 1 + \frac{B}{2A}$ and $C = \frac{\kappa}{\rho_{\rm{0}} c_{\rm{0}}^{4}} \left(\frac{1}{c_{\rm{v}}} - \frac{1}{c_{\rm{p}}} \right)$.

The following material constants of water were used in our simulations: the shear viscosity $\eta = 1.002 \times 10^{-3}$~Pa~s,  bulk viscosity $\zeta = 3.09 \times 10^{-3}$~Pa~s, specific heat capacity at constant pressure $c_{\rm{p}} = 4.182 \times 10^{3}$~J/(kg~K) and constant volume $c_{\rm{v}}  = c_{\rm{p}} /1.33$, and  thermal conductivity $\kappa = 0.597$~W/(m~K) \cite{tables}.

Eqs.~\ref{eq:A6}-\ref{eq:A7} are cumbersome but their solution by means of the standard FDTD procedure is straightforward and it does not suffer from long-term instabilities possible in some nonlinear FDTD schemes \cite{Mak11}. However, to achieve a better stability one needs to use a more stringent time step, as directed by the Courant stability factor $\Delta_{\rm{t}} = 0.5 \Delta_{\rm{z}}/c_{\rm{0,max}}$. Naturally, the standard Courant factor (without the $0.5$ pre-factor) works very well in the case of linearised equations, i.e. when $\beta = 0$.

The selection of the right value of $\Delta_{\rm{z}}$ is essential for conducting a high-accuracy simulation. (We assume that in the $2$D model the steps in space along the $x-$ and $y-$ directions are both equal to $\Delta_{\rm{z}}$.) Normally, a high accuracy is achievable by choosing $\Delta_{\rm{z}} < \lambda / 10$ where $\lambda$ is the shortest acoustic wavelength present in the model. For instance, for the acoustic frequency $1$ GHz and $\Delta_{\rm{z}} = 2$ nm one obtains $750$ spatial points per wavelength. Whereas a $50$ times larger step in space would already suffice to ensure a high accuracy, $\Delta_{\rm{z}} = 2$ nm is used in this paper because we also model the antenna whose cross-section equals $30$ nm, which is equivalent to $15$ spatial points. (Simulation of the antenna would be inaccurate if the cross-section were $\simeq 10$ spatial points).

Furthermore, we note that in the acoustic simulations the Courant stability criterion produces $\Delta_{\rm{t}} \approx 5 \times 10^{-13}$ s when $\Delta_{\rm{z}} = 2$ nm. It is also well-known that at least $3-4$ oscillation periods of the sinusoidal acoustic wave have to be simulated to obtain the correct result. To meet this condition, for the acoustic frequency $1$ GHz one needs to conduct $\sim 10^{4}$ iterations, which is equivalent to $\sim 15$ minutes of CPU time when the simulation is run on a multi-CPU workstation. Now let us assume that the acoustic frequency equals $1$ MHz. As we still use $\Delta_{\rm{z}} = 2$ nm required to correctly model the antenna, we obtain $750000$ spatial points per wavelength and we need to conduct $\sim 10^{7}$ iterations. This requires $\sim 2500$ hours of CPU time. Such a long simulation time is prohibitive even though a significant acceleration might be achieved by running the simulation on a supercomputer. This discussion additionally justifies our approach to model GHz-range hypersonic waves and project the results to the case of ultrasonic waves having lower frequencies.

\section{Connecting the acoustic and the optical FDTD models}

In this section we show how to connect the a-FDTD model with the o-FDTD model. We also discuss the nonlinear o-FDTD used to simulate the purely optical properties of the antenna. We start with the basic equations of the o-FDTD model:
\begin{align}
\begin{aligned}
- \mu_{\rm{0}} \frac{\partial \bm{H}(\bm{r},t)}{\partial t} = \nabla \times \bm{E}(\bm{r},t)
\end{aligned}
\label{eq:B1},
\end{align}
and
\begin{align}
\begin{aligned}
\epsilon  \epsilon_{\rm{0}} \frac{\partial \bm{E}(\bm{r},t)}{\partial t} + \bm{J}(\bm{r},t) = \nabla \times \bm{H}(\bm{r},t)
\end{aligned}
\label{eq:B2},
\end{align}
\noindent where $\bm{E}(\bm{r},t)$ is the electric field, $\bm{H}(\bm{r},t)$ is the magnetic field, and $\bm{J}(\bm{r},t)$ is the electric current density required to implement the optical Kerr nonlinearity \cite{Mak11}. The dielectric permittivity term $\epsilon$ also enters this equation and, as the optical properties of silver are modelled by using a Drude model \cite{sullivan}, in the absence of acoustic waves it can be assumed to be independent of $\bm{r}$.

However, in the presence of acoustic waves in agreement with Ref.~\onlinecite{Law01} the total permittivity $\epsilon_{\rm{tot}} (\bm{r},t) = \epsilon + \delta \epsilon (\bm{r},t)$ has to be introduced. The fluctuations of the dielectric permittivity due to the propagating acoustic waves $\delta \epsilon (\bm{r},t)$  is calculated from the values of the vector velocity field $\bm{v}(\bm{r},t)$ produced by the a-FDTD method. 

Consequently, Eq.~\ref{eq:B2} reads
\begin{align}
\begin{aligned}
\left[ \epsilon + \delta \epsilon (\bm{r},t) \right] \epsilon_{\rm{0}} \frac{\partial \bm{E}(\bm{r},t)}{\partial t} + \bm{J}(\bm{r},t) = \nabla \times \bm{H}(\bm{r},t)
\end{aligned}
\label{eq:B3}.
\end{align}

We remind that in the simulations of the plasmon-enhanced BLS we assume that outside the antenna $\delta \epsilon$ is independent of the space coordinate $\bm{r}$. However, in Eq.~\ref{eq:B3} we keep this dependence for the sake of generality.

The presence of $\delta \epsilon (\bm{r},t)$ in Eq.~\ref{eq:B3} is the only principal difference between the o-FDTD method used in this work to simulate the BLS effect and the standard electromagnetic FDTD method \cite{sullivan}, which also used in Section \ref{Design of the optical antenna} to investigate the purely optical properties of the antenna. Consequently, it is straightforward to derive the discrete analogues of Eqs.~\ref{eq:B1} and \ref{eq:B3}.

The iterative solution of the discrete equations is only possible by properly addressing the mismatch between the frequency of light and the frequency of the acoustic waves. As has been shown in the main text, the dielectric permittivity fluctuations are quasi-uniform at  frequencies below or equal to $1$ GHz. This observation drastically simplifies the stimulations. We additionally accelerate the simulations by taking advantage of a slow change of the dielectric permittivity as compared with the optical response of the antenna. We run the o-FDTD simulator at discrete values of time at which the dielectric permittivity change is large enough to modify the optical response.

The consideration of the nonlinear optical Kerr effect is also not straightforward because nonlinear effect are not taken into account in the standard o-FDTD formulations \cite{sullivan}. As shown in Ref.~\onlinecite{Mak11}, in the time domain the polarisation associated with the Kerr nonlinearity can be written as 

\begin{align}
\begin{aligned}
\bm{P}_{\rm{Kerr}} (t) = \epsilon_{\rm{0}} \chi^{(3)} |\bm{E}(t)|^{2} \bm{E}(t)
\end{aligned}
\label{eq:B4}.
\end{align} 

\noindent We use one of the noniterative updating schemes presented in Ref.~\onlinecite{Mak11}. To fulfil the causality principle of nonlinear optics \cite{Boyd}, we discretise Eq.~\ref{eq:B4} as 

\begin{align}
\begin{aligned}
\bm{P}_{\rm{Kerr}} (t_{\rm{n+1}}) = \epsilon_{\rm{0}} \chi^{(3)} |\bm{E}(t_{\rm{n}})|^{2} \bm{E}(t_{\rm{n+1}})
\end{aligned}
\label{eq:B5},
\end{align} 

\noindent where $n$ stands for the time in step used in the o-FDTD model and $\chi^{(3)}$ is the third-order nonlinear susceptibility of the medium. Then we substitute Eq.~\ref{eq:B5} into the constitutive equation $\bm{D}(\bm{r},t) = \epsilon \epsilon_{\rm{0}} \bm{E}(\bm{r},t) + \bm{P}_{\rm{Kerr}} (\bm{r},t)$, where we neglected the dependence of the dielectric permittivity $\epsilon$ on the fluctuations caused by the acoustic waves because these fluctuations are irrelevant to the study of the nonlinear optical properties of the antenna. Finally, we obtain the following noniterative electric-field updating equation 

\begin{align}
\begin{aligned}
E_{\rm{q}}^{\rm{n+1}} = D_{\rm{q}}^{\rm{n+1}} / \left(\epsilon \epsilon_{\rm{0}} + \epsilon_{\rm{0}} \chi^{(3)} |\bm{E}^{\rm{n}}|^{2} \right)
\end{aligned}
\label{eq:B6},
\end{align} 

\noindent where $q=\langle x, y, z \rangle$ denotes the components of the electric field vector $\bm{E}$ in the Cartesian coordinate system and the discrete time $t_{\rm{n}}$ has been replaced by the index $n$.


\end{document}